\documentstyle[12pt,aaspp4,epsf]{article}
\def\fun#1#2{\lower3.6pt\vbox{\baselineskip0pt\lineskip.9pt
  \ialign{$\mathsurround=0pt#1\hfil##\hfil$\crcr#2\crcr\sim\crcr}}}
\def\lap{\mathrel{\mathpalette\fun <}}
\def\gap{\mathrel{\mathpalette\fun >}}

\def\etal{{\it et al.}}

\slugcomment{{\bf Submitted to The Astrophysical Journal}, January 1997}

\lefthead{D. Merritt}
\righthead{Cusps and Triaxiality}

\begin{document}

\title{Cusps and Triaxiality}

\author{David Merritt}
\affil{Department of Physics and Astronomy, Rutgers University,
New Brunswick, NJ 08855}

\begin{abstract}
Statler (1987) demonstrated that self-consistent triaxial models 
with the perfect density law could be constructed for virtually
any choice of axis ratios.
His experiments are repeated here using triaxial mass models based on 
Jaffe's density law, which has a central density that diverges as 
$r^{-2}$, similar to what is observed in low-luminosity elliptical 
galaxies.
Most of the boxlike orbits are found to be stochastic in these
models.
Because timescales for chaotic mixing are generally shorter than a 
galaxy lifetime in triaxial models with strong cusps, and because 
fully-mixed stochastic orbits have shapes that are poorly suited 
to reproducing a triaxial figure, only the regular orbits are
included when searching for self-consistent solutions.
As a result of the restriction to regular orbits, self-consistent
solutions are found only for mass models with a modest range of 
shapes, either nearly oblate, nearly prolate or nearly spherical.
This result may explain in part the narrow range of elliptical 
galaxy properties.
\end{abstract}

\section {Introduction}

The hypothesis that elliptical galaxies are generically triaxial  
in shape has been challenged from two directions in recent years.
Observational evidence for triaxiality, which at one time seemed 
compelling, is now viewed with more skepticism.
While isophotal twists are fairly common in early-type 
galaxies, many of these systems are probably barred S0's,
not bona-fide ellipticals (\cite{fab89}; \cite{nib92}).
Kinematical tests for triaxiality, when applied to uniformly 
selected samples, suggest that few if any elliptical galaxies
need be strongly triaxial in shape (\cite{fri91}).
Detailed modelling of a few ellipticals with 
well-observed rotational velocity fields has mostly failed to 
reveal convincing evidence for triaxiality, at least in the central 
regions where the stellar distribution is likely to be
relaxed (\cite{sta94}; \cite{qid95}; \cite{sts96}).
Although strong departures from axisymmetry are observed in a few 
bona-fide elliptical galaxies, many of these have probably undergone tidal 
interactions or mergers in the recent past and hence are not in a 
steady state (e.g. NGC 1549, \cite{lon94}; NGC 4589, \cite{mob89}; 
Centaurus A, \cite{hui95}).
And the discovery of lenticular systems with significant 
counterrotation (\cite{rug92}; \cite{mek92}) has weakened the conceptual 
link between velocity anisotropy and triaxiality.

On the theoretical side, a number of recent studies have emphasized the 
predominance of chaotic motion in the phase space of realistic 
triaxial potentials (\cite{sch93}; \cite{mef96}; \cite{mev96};
\cite{pal97}).
Most of the boxlike orbits in triaxial models with strong cusps, 
i.e. models in which the central density diverges more rapidly 
than $\rho\propto r^{-1}$ at the center, are stochastic.
Such cusps appear to be ubiquitous in low-luminosity ellipticals,
$M_V\gap-20.5$ (\cite{geb96}).
Furthermore, numerical experiments demonstrate that the chaos 
in triaxial models with high central densities
can redistribute stars in phase space on timescales that are 
short compared to galaxy lifetimes (\cite{mev96}).
The mechanism that drives this relaxation is chaotic mixing 
(\cite{kam94}, 1995); in a fully-mixed system, the phase space density is 
constant throughout the stochastic parts of phase space at every energy.
The loss of freedom that results from forcing the stochastic 
orbits to be fully mixed can preclude a self-consistent 
triaxial equilibrium (\cite{mef96}).

The goal of the present paper is to determine the degree to which
triaxiality can be supported in galaxies with high central 
densities.
The approach is similar to that of Statler (1987), who 
constructed self-consistent models of triaxial galaxies with the 
``perfect'' density law, characterized by an unphysically large core.
Orbits in the perfect ellipsoid are all regular and fall into one 
of just four families: the short-axis tubes; the 
inner- and outer long-axis tubes; and the boxes (\cite{kuz73}; 
\cite{del85}).
Statler showed that linear superposition of orbits from the four 
families could reproduce the mass distribution of the perfect ellipsoid
for virtually any choice of the axis ratios $c/a$ and $b/a$.
Furthermore he found that his solutions were highly nonunique, 
in the sense that stars could be apportioned in significantly 
different ways between the four families without violating 
self-consistency.
However all of his solutions required significant numbers of 
stars to be assigned to box orbits.
The importance of box orbits for the triaxial self-consistency problem
was stressed also by Schwarzschild (1979, 1982) who constructed models 
based on the modified Hubble law.
While orbits in the triaxial Hubble model are not all regular,
Schwarzschild's models contained only small numbers of stochastic orbits 
(\cite{mer80}).

The absence of box orbits in triaxial models with 
realistic density profiles might be expected to limit the 
allowed degree of triaxiality.
Schwarzschild (1993) found this to be true in highly flattened, scale-free 
($\rho\propto r^{-2}$) triaxial models, and Merritt \& Fridman 
(1996, hereafter Paper I) showed
that a ``maximally triaxial'' model with Jaffe's (1983) density 
law
\begin{equation}
\rho(m)=\rho_0 m^{-2}(1+m)^{-2}, \ \ \ m^2={x^2\over a^2}+{y^2\over 
b^2}+{z^2\over c^2}
\end{equation}
and $c/a=0.5$ could not be reproduced using just the regular 
orbits.

Here, the numerical experiments of Paper I are extended to 
triaxial mass models with a range of axis ratios. 
Jaffe's law (1) for the radial density profile is adopted 
throughout;
thus these models are reasonable representations of elliptical galaxies 
fainter than $M_V\approx -20.5$, almost all of which are observed to 
contain central density cusps with logarithmic slopes $\gamma$ close 
to $2$ (\cite{geb96}).
Brighter ellipticals, $-22\lap M_V\lap -20.5$, also have power-law cusps 
but with a range of slopes, $0\lap\gamma\lap 2$; the 
brightest ellipticals, $M_V\lap -22$, tend to have the weakest cusps, 
$\gamma\lap 1$ (\cite{mef95}).
There is increasingly strong evidence for nuclear black 
holes containing $\sim 1\%$ of the stellar mass in elliptical 
galaxies of all luminosities (\cite{kor95}).
Since the behavior of boxlike orbits in triaxial models with 
central singularities is similar to that of models with strong 
cusps (\cite{mev96}), the models studied here are likely to be 
relevant at some level to elliptical galaxies of all luminosities.

Libraries of orbits were computed in the potentials corresponding to
these mass models, and linear superpositions of orbits were sought 
that yielded the mass of the model in a discrete grid of cells.
Orbits were tested for stochasticity by computing their Liapunov 
exponents; as expected, most of the boxlike orbits were found to be 
stochastic.
In a strictly time-independent model, stochastic orbits can be 
included only in the form of invariant ensembles, one per energy, 
that correspond to a uniform filling of chaotic phase space
(\cite{kam94}).
Such ensembles are generally rounder than the model and not very useful for 
reconstructing the assumed mass distribution.
Furthermore, in triaxial potentials with strong cusps, the 
chaotic mixing timescale over which stochastic phase space 
would be uniformly filled is generally 
shorter than a galaxy lifetime, particularly near the galaxy 
center (\cite{mev96}).
Hence, in the current study, the stochastic orbits were simply 
deleted from the orbit libraries when searching for self-consistent 
solutions.

This limitation to regular orbits -- mostly tubes, with some 
``boxlet'' families -- was found to 
impose severe restrictions on the shapes of self-consistent
models.
Strongly triaxial models, which would heavily weight the box orbits, 
are ruled out.
Only nearly oblate, prolate or spherical models are allowed;
such models require only a small contribution from non-tube orbits.
These results suggest that the 
striking regularity in the observed properties of elliptical 
galaxies -- the avoidance of strongly triaxial shapes, and the
narrow range of kinematical behaviors -- might be due in part 
to the restrictions imposed by chaos on the population of phase space.

The algorithms used here to compute orbits, evaluate Liapunov exponents 
and construct self-consistent solutions were similar in most 
respects to those described in Paper I.
Accordingly, only a brief discussion of the numerical methods is
given here in \S2.
\S3 presents the grid of 25 triaxial mass models and their orbital 
populations.
The results from the self-consistent modelling are presented in 
\S4.
Implications for the structure of real elliptical galaxies 
are discussed in \S5.

\section{Numerical Methods}

The grid of 25 mass models is displayed in Figure 1.
The short-to-long axis ratio $c/a$ was assigned a value 
from the set $(0.4,0.5,0.6,0.7,0.8)$, and the triaxiality parameter 
$T=(a^2-b^2)/(a^2-c^2)$ was set equal to $0.1$ (nearly oblate), 
$0.3,0.5,0.7,$ or $0.9$ (nearly prolate).

For each mass model, 6840 orbits were integrated for 50 dynamical 
times $T_D(E)$, defined as the period of the 1:1 resonant orbit 
in the $x-y$ plane at energy $E$.
A 7/8 order Runge-Kutta algorithm with variable step size was used
for the integrations.
Initial conditions were assigned as in Paper I, from one of two
grids of starting points: either on an equipotential surface with 
zero initial velocity (stationary), or in the $x-z$ plane with 
$v_x=v_z=0$ ($x-z$).
Orbits started in the $x-z$ plane are mostly tube orbits, while 
those started on an equipotential surface are boxlike, passing 
close to the center.
Orbits with both sets of initial conditions were assigned one of 
a set of 20 energies, defined as the values of the potential on 
the $x$-axis of a set of ellipsoidal shells -- with the same axis 
ratios as the density -- that divide the model into 21 sections 
of equal mass.
The grid of starting points at each energy is described in Paper 
I; it contained 150 orbits per shell in $x-z$ initial condition 
space, and 192 orbits per shell in stationary initial condition 
space.

Stochasticity was detected by computing the largest Liapunov exponent 
$\sigma_1$ for each orbit, defined as the average (over 50 
dynamical times) of the exponential rate of divergence between 
the orbit and an infinitesimally-nearby orbit.
At each energy and for each of the initial condition grids,
a histogram of the $\sigma_1$ values was constructed.
These histograms invariably exhibited strong bimodality, with 
one narrow peak near zero (the regular orbits) and another, more 
diffuse peak centered around larger values of $\sigma_1$ 
(the stochastic orbits).
As in Paper I, this scheme might sometimes mistake 
stochastic orbits for regular orbits, since some stochastic orbits 
remain ``trapped'' near regular parts of phase space for long 
periods of time.
However, any stochastic orbits that are mis-classified in this 
way are effectively regular, since they can be counted on to mimic 
regular orbits for many dynamical times.

\section {Orbit Families}

Figure 2 gives the average number of stochastic orbits per shell 
in both initial-condition spaces for each of the 25 mass models.
Typically only a few percent of the orbits from the $x-z$ initial 
condition space were found to be stochastic, but the fraction of 
stochastic orbits from the stationary initial condition space was much 
larger, usually more than one-half.
While the numbers in Figure 2 can not be simply translated 
into phase-space fractions, they suggest that stochasticity of 
the boxlike orbits is most important in strongly triaxial or nearly 
prolate mass models, and least important in nearly oblate models.
(Of course, the boxlike orbits occupy only a small fraction of 
the total phase space in nearly-axisymmetric models.)
Many workers have noted the ``triaxial'' nature of the 
orbit populations in nearly-prolate models.

Figures 3 and 4 illustrate the two initial condition spaces at 
shell 10 for 12 of the mass models.
Regular orbits were assigned to one of three families: long-axis 
tubes, short-axis tubes and boxes.
Long-axis tubes have a nonzero averaged 
angular momentum about the $x$-axis, short-axis tubes have a 
definite $L_z$, and boxes have no obvious, time-averaged 
angular momentum components.

The starting points of the three most important resonant orbit 
families are also plotted in Figures 3 and 4.
The 1:1 orbits in the $x-y$ plane generate most of the short-axis 
tubes; the 1:1 orbits in the $y-z$ plane generate the long-axis 
tubes; and the $1:2$ $x-z$ ``banana'' orbits generate many 
of the regular boxlike orbits.
As found by Schwarzschild (1993), the number of minor resonances 
that generate regular orbits in the stationary initial condition 
space increases as the model becomes rounder.

\section{Self-Consistent Models}

The fraction of time spent by each orbit in a grid of cells was 
recorded, and a linear superposition of orbits was sought, with 
non-negative occupation numbers, that reproduced the known mass 
of the model in each cell.
The details of the solution grid and of the optimization routine 
are given in Paper I.

When the full set of 6840 orbits was supplied to the optimization 
routine, self-consistent solutions were found for almost 
all of the 25 mass models. 
However, as discussed in Paper I, such solutions are not 
bona-fide equilibria, since the optimization routine is free to 
assign arbitrary occupation numbers to the 
different stochastic orbits at each energy.
In a real galaxy, chaotic mixing would tend to produce a uniform 
population of stochastic phase space at each energy on a 
timescale of order $10^2$ dynamical times (\cite{mev96}).
Thus, these quasi-equilibrium solutions would be expected to
slowly evolve.

Fully stationary models can be constructed in one of two ways: by 
eliminating the stochastic orbits from the orbit libraries; or by 
replacing the set of stochastic orbits at each energy by an 
invariant ensemble representing a uniform population of 
stochastic phase space.
The latter procedure was implemented in Paper I, where it was
found that the inclusion of invariant ensembles 
reduced slightly the the residuals of the best-fit orbit 
population below those of a solution containing regular 
orbits alone.
Here, the simpler alternative of omitting the 
stochastic orbits was chosen.
This choice is not meant to suggest that nature would 
avoid placing stars on stochastic orbits in a real galaxy.
However the invariant ensembles -- because of their generally 
spherical shapes -- are only a minor asset from the 
point of view of reconstructing the triaxial figure, and by 
omitting them one is likely to restrict only slightly the range 
of shapes that can be reproduced in a fully stationary way.

(This argument would be much less valid if applied to triaxial models 
with weak cusps or cores, in which the chaotic mixing timescales
can be much longer (\cite{gos81}; \cite{mev96}; Paper I).
In such systems, a highly nonuniform population of stochastic 
phase space could presumably be maintained for a long period of 
time -- in effect, many different stochastic orbits would exist 
at the same energy, giving much more freedom to construct a 
self-consistent equilibrium.
Imposing secure limits on the degree of triaxiality would be 
difficult for such models since the range of allowable orbital 
populations would depend strongly on the degree to which 
stochastic phase space was assumed to be mixed.
Most of this mixing presumably takes place during galaxy 
formation and so a solution to the self-consistency problem would 
require some knowledge about the formation process.)

Mass models for which self-consistent solutions containing only 
regular orbits were found are delineated in Figure 1.
Only nearly axisymmetric, or spherical, mass models could be 
reconstructed from the regular orbits alone.
Roughly speaking, models with $T\lap 0.4$, $T\gap 0.9$ or $c/a\gap 0.8$
were found to have such solutions; the range in allowed values of $T$ 
decreases with decreasing $c/a$, and no solutions containing just regular orbits 
were found for $c/a=0.4$.

The fraction of the mass which these solutions place on the 
three orbit families are indicated in Figure 5.
Bold-faced numbers refer to models for which self-consistency was 
achieved; in the other models, the mass fractions shown in Figure 
5 are those corresponding to the orbital population that most nearly 
reproduced the mass of the model in the cells.
The variation found here in the orbital populations over the $(c/a,T)$ plane 
may be compared to that found by Statler (1987, Fig. 6) in his study of 
fully integrable triaxial models.
Nearly oblate (prolate), self-consistent models were found to contain a 
preponderance of short (long) axis tubes, similar to the results of Statler, 
and consistent with the expected form of the solutions in the 
axisymmetric limits.
However the fraction of box orbits was found to be small, $\lap 
20\%$, in all of the self-consistent solutions found here, even the 
substantially triaxial ones.
By contrast, Statler found a much larger mass fraction in boxes, 
typically between 50\% and 75\% in models with $c/a\lap 0.6$.
This difference is presumably due to the much narrower range of 
regular box orbit shapes in the potentials considered here, which 
makes them less useful for reproducing the figure.
The largest box-orbit fractions found here were in models for 
which self-consistency could not be achieved; however even in these 
failed solutions the box orbit fraction never exceeded 40\%.

Statler (1987) stressed the non-uniqueness of his solutions, i.e. 
the considerable freedom which he found to shift stars from one 
orbit family to another without violating self-consistency.
One would expect much less degeneracy in our solutions, especially 
those lying near the curve in Figure 1 that divides allowed shapes 
from forbidden ones.
Along this curve, which delineates the edge of solution 
space, the orbital populations are probably unique.
The degree of non-uniqueness is likely to increase toward the 
oblate and prolate axes, since axisymmetric models are known to 
be degenerate in their orbital populations (e.g. \cite{deg93}).
However no attempt was made here to explore the degree of degeneracy of 
the solutions.

\section {Discussion}

A long-standing puzzle in the study of elliptical 
galaxies is the narrow range of morphological and 
kinematical properties which they exhibit.
Kinematical tests for non-axisymmetry in well-selected samples 
find little evidence for strong triaxiality; most ellipticals 
appear to be either nearly oblate or nearly prolate 
(\cite{fri91}).
Detailed studies of the velocity fields of individual galaxies 
also imply that axisymmetry is the norm, at least in galaxies
where the stellar distribution is likely to be relaxed 
(\cite{sta94}; \cite{qid95}; \cite{sts96}).
Among elliptical galaxies of a given luminosity, 
the distribution of Hubble types is quite narrow.
Bright ellipticals have an apparent shape 
distribution that is peaked at $b/a\approx 0.85$ 
with a dispersion of only $\sim 0.1$ (\cite{ryl93}).
This distribution is inconsistent with complete axisymmetry but only 
mildly so (\cite{trm96}). 
The Hubble-type distribution of fainter ellipticals is not so 
well determined but is also fairly narrow; the peak lies at 
an apparent axis ratio of $\sim 0.7$ and the distribution is 
fully consistent with axisymmetry (\cite{trm96}).

In their kinematics, too, elliptical galaxies show surprisingly
little variation.
The line-of-sight velocity distributions, whose shapes 
reflect to some degree the average character of the stellar orbits, 
are almost always nearly Gaussian.
The most prominent deviations are asymmetric ones resulting 
from the rotation of a distinct subcomponent (e.g. \cite{fri88}); 
large symmetric deviations -- like those
expected in spherical models with strong velocity anisotropy, for 
instance -- are almost never observed (\cite{bes94}).
Together with the virial theorem, this homogeneity in the 
shapes and kinematics of elliptical galaxies implies that their  
observable scaling parameters (diameter, velocity dispersion, surface 
brightness) should be tightly correlated.
This is in fact the case: elliptical galaxies define a 
``Fundamental Plane'' which is extremely thin (\cite{djd87}).

Until recently, theoretical work provided little in the way of 
explanation for this homogeneity.
Studies of the axisymmetric and triaxial self-consistency 
problems (mostly in the context of mass models with large cores)
demonstrated again and again the extremely wide 
range of possible solutions, both morphological and kinematical 
(\cite{sta87}; \cite{deg93}; \cite{hun95}).
Stability studies, while ruling out some models extreme in their 
shapes (e.g. \cite{mes94}) or kinematics (e.g. \cite{sah91}),
failed to narrow this range appreciably.
$N$-body simulations of elliptical galaxy formation also tended to 
generate models with a much wider range of properties than observed.
The shapes and kinematics of galaxies produced by the merger of 
$N$-body disks, for instance, are strongly correlated 
with the relative orientation of the colliding galaxies (\cite{bar92}).
These $N$-body remnants are often highly flattened, triaxial, 
or box-shaped, in sharp contrast with the majority of observed 
ellipticals.
Dark halos produced in simulations of hierarchical structure 
formation tend to be highly elongated and triaxial
(\cite{war92}), again quite unlike real elliptical galaxies. 

We propose here that the narrow range of elliptical galaxy 
shapes and kinematics may be a simple consequence of 
dynamical self-consistency.
We consider first faint ellipticals, which have stellar density 
cusps that 
are predictably as steep as $r^{-2}$.
Figure 1 shows that stationary models with strong cusps 
are very limited in their allowed shapes: they must be nearly oblate
($T\lap 0.4$), nearly prolate ($T\gap 0.9$), or nearly spherical 
($c/a\gap 0.8$).
Given the narrowness of the allowed region along the 
prolate boundary in Figure 1, we might further expect 
oblate and spherical shapes to be preferred over prolate ones.
These predictions are consistent with what little is known about
the intrinsic shapes of low-luminosity elliptical galaxies.
The flattenings of faint ellipticals correlate with their 
rotation in the manner expected for ``oblate isotropic rotators''
(\cite{dav83}), circumstantial evidence that they are 
approximately oblate.
Axisymmetry is also consistent with the frequency function of 
Hubble types of faint ellipticals, but the same is true of
a triaxial shape distribution (\cite{trm96}).
Detailed modelling of two-dimensional velocity fields has now 
been carried out for a few galaxies; among these, M32 is faint,
$M_V=-16.3$, and appears to be accurately oblate (\cite{qid95}).

Our results might also predict a narrow range of {\it kinematical} 
properties for galaxies with strong cusps.
Suppose that nature constructed a galaxy with a strong cusp and
with an initial shape that placed it outside of the region corresponding to 
self-consistent solutions in Figure 1.
Such a galaxy would not be able to reach an equilibrium state 
without changing its shape.
Although the direction of that change is not obvious, the galaxy
would eventually cross the curve that separates allowed from
forbidden shapes in Figure 1.
Since the orbital population of an equilibrium model near 
this curve is likely to be unique -- as argued above -- one 
would expect galaxies so produced to have kinematical properties 
that are strongly correlated with their shapes.

(The orbital compositions of real galaxies produced in this way would 
probably differ somewhat from those of the models constructed 
here, since nature would not exclude stars from the stochastic 
parts of phase space.
However the mass fractions associated with stochastic orbits in 
fully stationary galaxies are likely to be modest (Paper I).)

The results presented here are not so clear in their implications for bright
elliptical galaxies, $M_V\lap-20.5$, which exhibit a range of cusp 
density slopes (\cite{mef95}).
Chaotic mixing timescales of boxlike orbits 
can be quite long when the cusp is weak or 
nonexistent (\cite{gos81}; \cite{mev96}), implying more 
freedom to reconstruct a triaxial shape.
However many elliptical galaxies, both bright and faint, may 
contain massive nuclear black holes like those in M32 and M87
(\cite{kor95}); in fact it has been suggested that the weak 
stellar cusps of bright ellipticals are produced during the 
coalescence of two black holes following a merger (\cite{ebi91}).
The properties of stochastic orbits in triaxial 
potentials with massive central singularities are similar to 
those in triaxial potentials with strong cusps (\cite{mev96}).
If massive nuclear black holes are ubiquitous, we might expect bright
ellipticals to also be strongly constrained in their shapes.
Interestingly, the Hubble-type distribution of elliptical galaxies with 
$M_V\lap -20.5$ is strikingly different from that of fainter 
ellipticals and appears to require a certain number of non-axisymmetric
galaxies (\cite{trm96}).
The sample of (mostly bright) ellipticals analyzed by Franx, Illingworth 
\& de Zeeuw (1991) also contains a few members with significant 
minor axis rotation, hence probably triaxial.
But these facts are not obviously inconsistent with Figure 1, 
since strongly triaxial shapes are permitted when $c/a\gap 
0.8$.
A large fraction of bright elliptical galaxies satisfy this 
condition (\cite{trm96}).

Many of the systematic differences between bright and faint 
ellipticals -- including the steepness of the cusps --
are thought to result from the greater importance of 
gaseous dissipation during the formation of the latter (\cite{fab97}).
Are the purely stellar-dynamical arguments presented here relevant 
to galaxies that formed dissipatively?
Barnes (1996) has discussed how the addition of a dissipative 
component to $N$-body simulations of disk galaxy mergers can 
affect the final shape and kinematics of the remnant.
In the absence of gas, the remnants are often strongly 
triaxial, containing many stars on box orbits.
When gas is included, gravitational torques during the merger 
deposit much of the gas near the center on a short timescale,
deepening the potential well there.
The stellar component of the galaxy responds by becoming 
more oblate.
Barnes \& Hernquist (1996) speculate that this change in shape is
driven by the destabilization of box orbits which pass near  
the center -- the same mechanism that is responsible for ruling out 
strongly triaxial models in the present study.
A number of authors have noted similar changes in the 
shape of an initially triaxial, $N$-body galaxy following an increase in the 
central density (e.g. \cite{nom85}; \cite{udr93}; \cite{dub93}).
The dissipative formation of galaxies is undoubtedly complex, 
but these arguments suggest that the evolution toward more oblate shapes
seen in numerical simulations with dissipation can be explained
in part by the absence of strongly triaxial equilibria.

\section{Summary}
Strong triaxiality is inconsistent with a high central density.
Dynamical models with Jaffe's density law, which has 
$\rho\propto r^{-2}$ near the center, must be nearly 
oblate ($T=(a^2-b^2)/(a^2-c^2) \lap 0.4$), nearly prolate ($T\gap 
0.9$) or nearly spherical ($c/a\gap 0.8$).
This result may explain in part the homogeneity of elliptical 
galaxies, including the tendency of low-luminosity 
ellipticals to be oblate, and the narrow range of elliptical galaxy 
kinematical properties.

\bigskip

This work was supported by NSF grant AST 90-16515 and by NASA 
grant NAG 5-2803.
G. Quinlan's assistance in the programming is gratefully acknowledged.
T. Fridman and M. Valluri kindly read a first draft and made 
useful suggestions for changes.

\clearpage

\setcounter{figure}{0}

\begin{figure}
\plotone{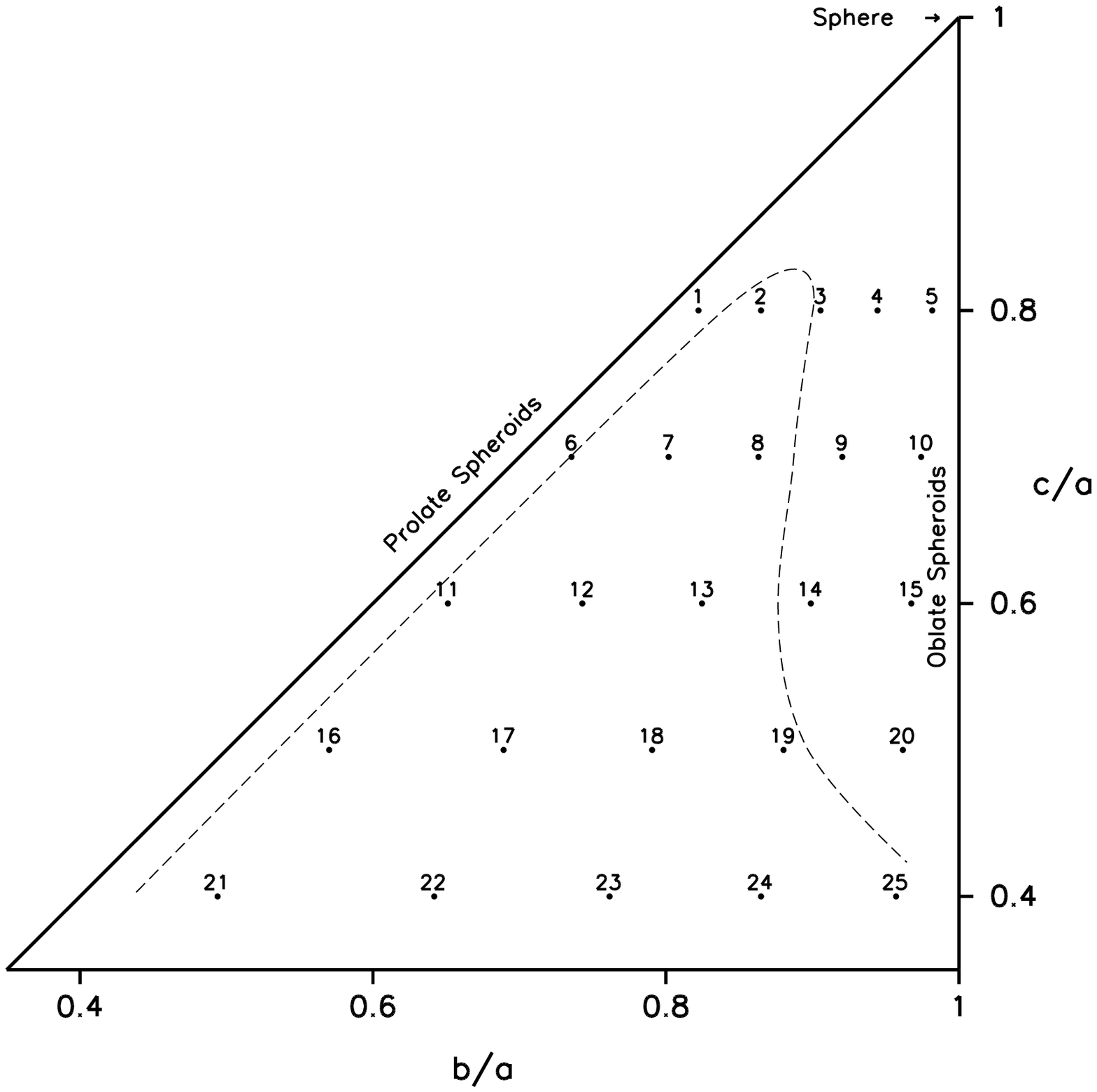}
\caption{Plane of axis ratios, showing
the 25 models of the survey.
The dotted curve denotes the approximate limit of solution space;
points below correspond to models for which no self-consistent
solution was found.}
\end{figure}

\begin{figure}
\plotone{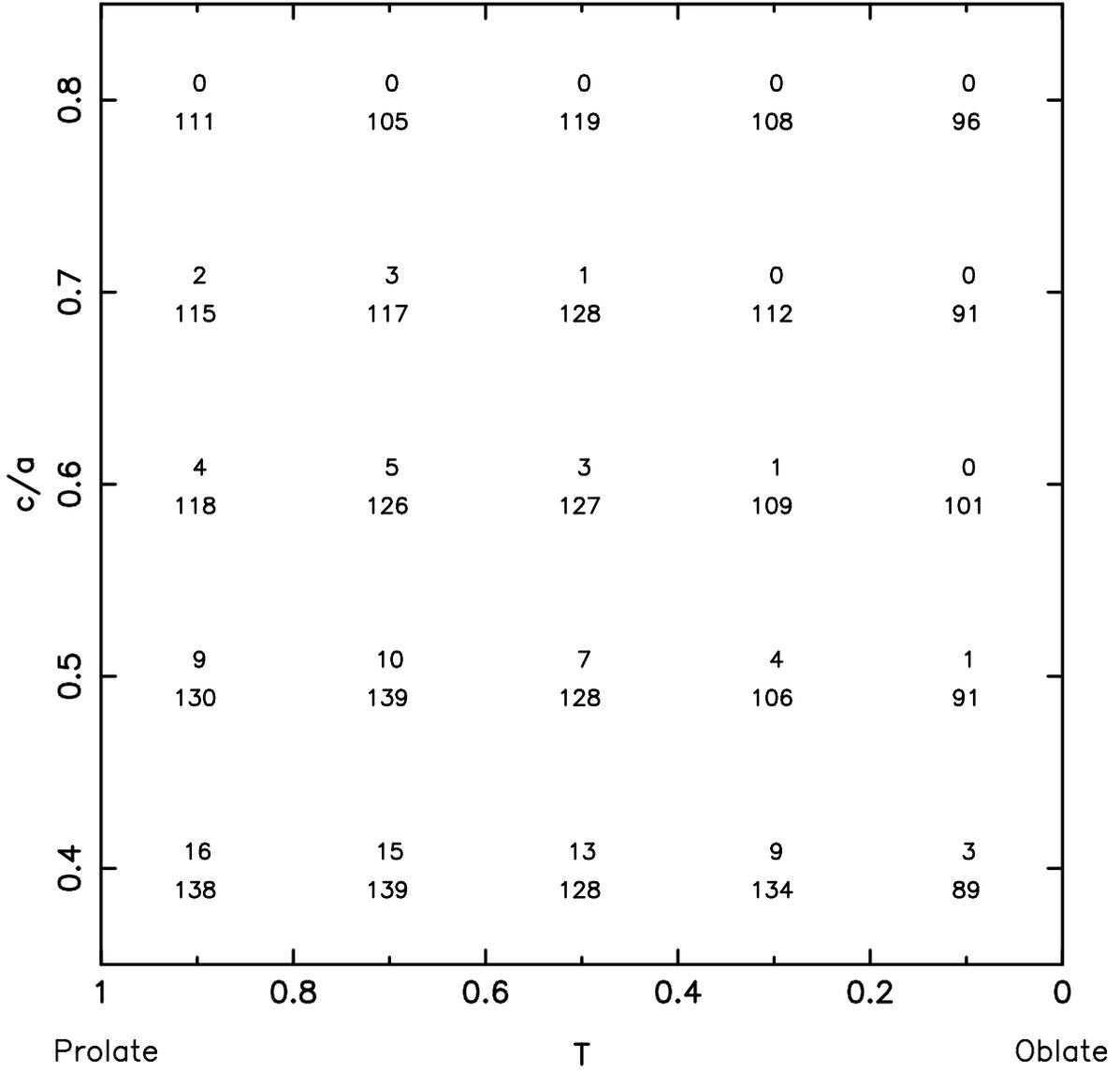}
\caption{Average number of stochastic
orbits per shell in the 25 models, out of a total of 150 ($x-z$ initial 
condition space; top numbers) or 192 (stationary initial condition space;
bottom numbers).}
\end{figure}

\begin{figure}
\plotone{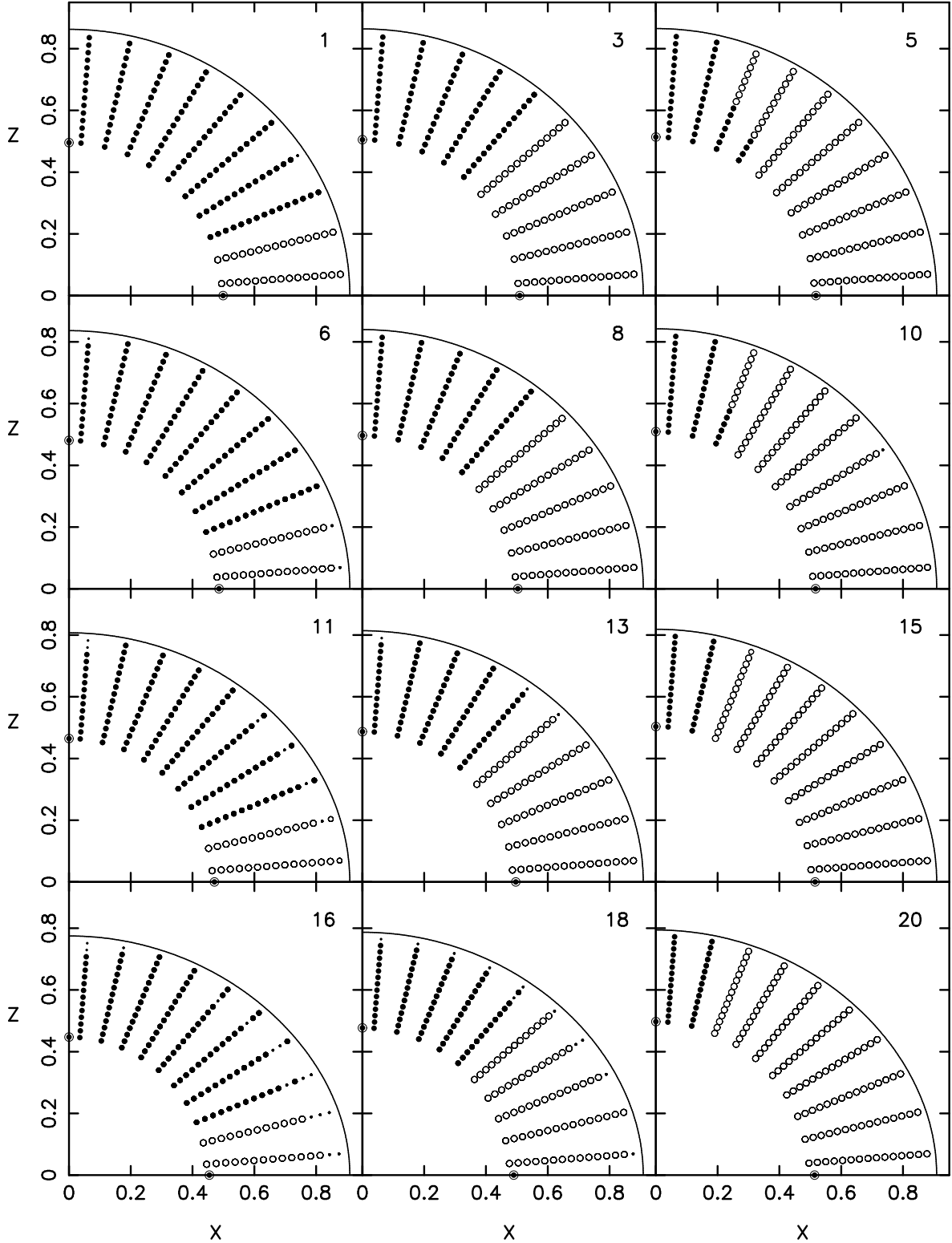}
\caption{$x-z$ initial condition space
at shell 10 for a set of models. 
Numbers in the upper right of each frame correspond to positions
in the axis-ratio plane of Figure 1.
Filled circles denote the long-axis tubes, large open circles the 
short-axis tubes, small open circles the boxes, and small dots 
the stochastic orbits.
Circled dots are the 1:1 resonant orbits in the $x-y$ and $y-z$ 
planes.}
\end{figure}

\begin{figure}
\plotone{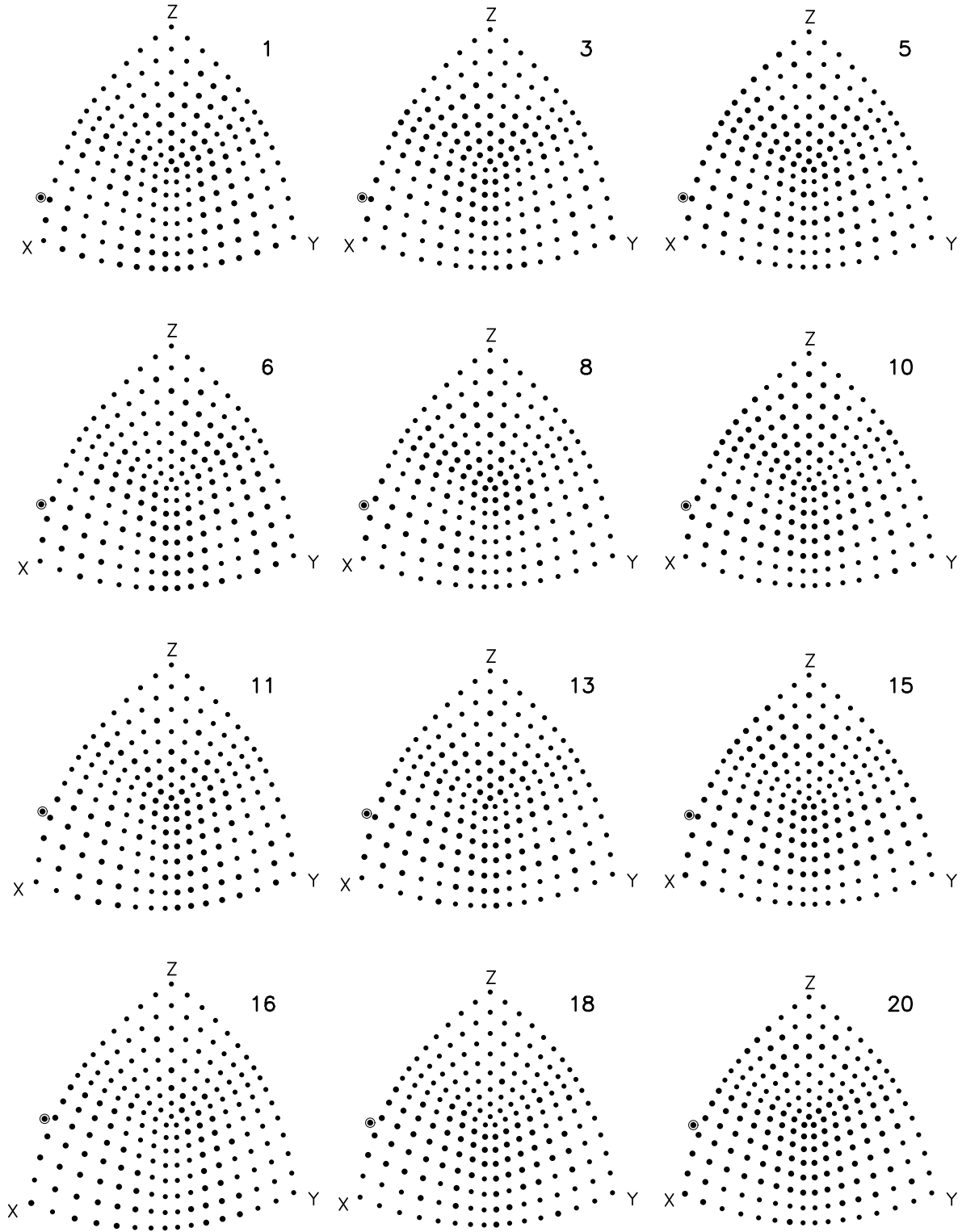}
\caption{Stationary initial condition space at shell 10 for a set of models. 
Numbers in the upper right of each frame correspond to the positions
in the axis-ratio plane of Figure 1.
Large dots denote the regular orbits and small dots 
the stochastic orbits.
Circled dots are the 1:2 resonant ``banana'' orbits in the $x-z$ plane.}
\end{figure}

\begin{figure}
\plotone{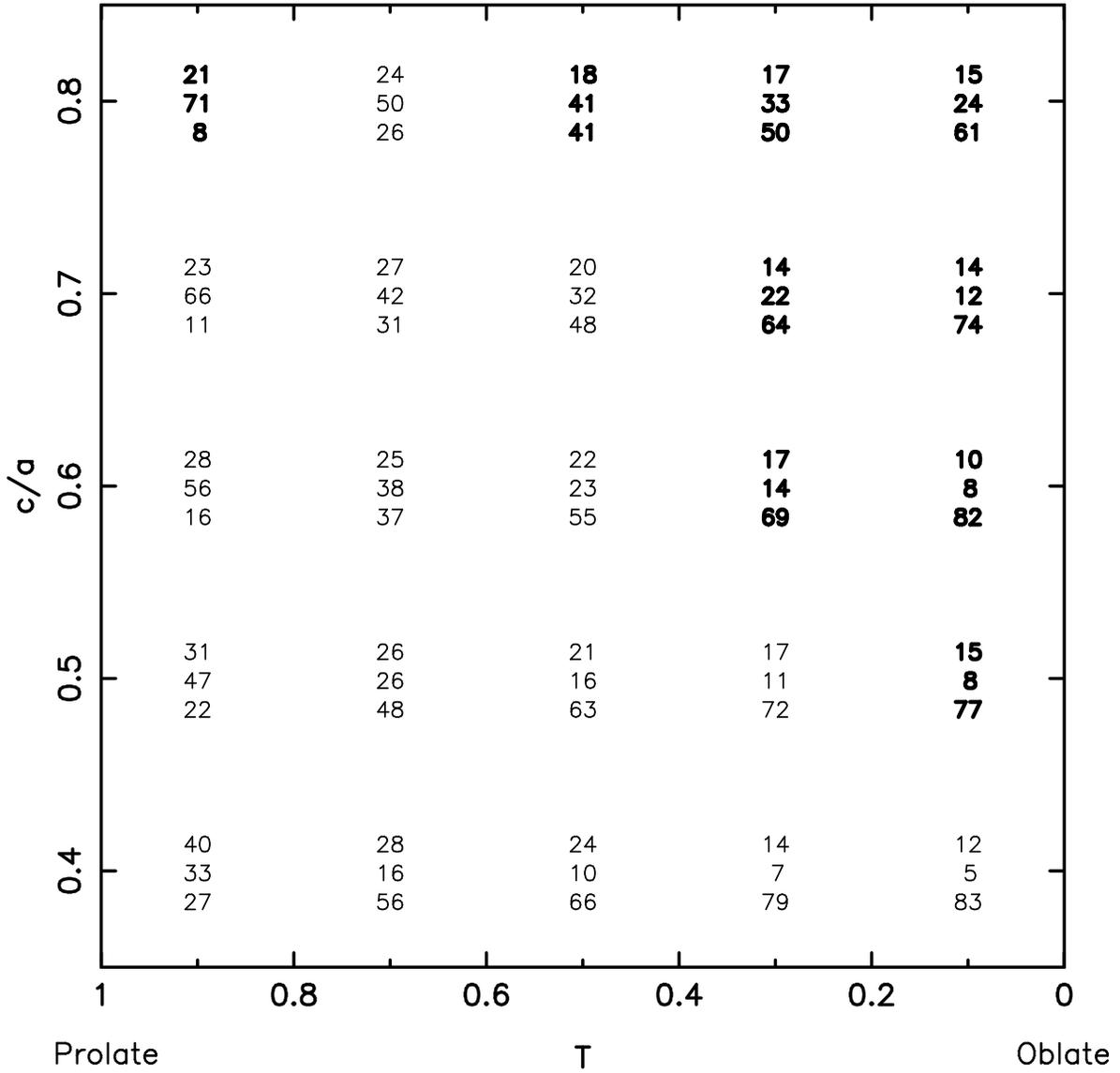}
\caption{Orbital mass fractions for 
the numerical solutions.
The top number indicates the percentage of the total mass in box 
orbits, the middle number the percentage in long-axis tubes, and the 
lower number the percentage in short-axis tubes.
Bold-faced numbers refer to models for which self-consistency was 
achieved.}
\end{figure}

\end{document}